# Measurements of photoelectron extraction efficiency from CsI into mixtures of Ne with $CH_4$, $CF_4$, $CO_2$ and $N_2$

**J. Escada,[a*] L.C.C. Coelho,[a] T.H.V.T. Dias,[a] J.A.M. Lopes,[a,b] J.M.F. dos Santos[a] and A. Breskin[c]**

[a] *GIAN, Departamento de Física, Faculdade de Ciências e Tecnologia, Universidade de Coimbra, Rua Larga, 3004-516 Coimbra, Portugal*

[b] *Instituto Superior de Engenharia de Coimbra, 3030-199 Coimbra, Portugal*

[c] *Department of Particle Physics, Weizmann Institute of Science, 76100, Rehovot, Israel*

*E-mail*: jescada@gian.fis.uc.pt

ABSTRACT: Experimental measurements of the extraction efficiency *f* of the UV-induced photoelectrons emitted from a CsI photocathode into gas mixtures of Ne with $CH_4$, $CF_4$, $CO_2$ and $N_2$ are presented; they are compared with model-simulation results. Backscattering of low-energy photoelectrons emitted into noble gas is significantly reduced by the admixture of molecular gases, with direct impact on the effective quantum efficiency. Data are provided on the dependence of *f* on the type and concentration of the molecular gas in the mixtures and on the electric field.

[*] Corresponding author.

## Contents

## 1. Introduction

CsI UV-photocathodes [1] are routinely used in gas photomultipliers (GPMTs) with many gas-avalanche electron-multiplier configurations, mostly to detect Cherenkov radiation-induced primary or/and secondary scintillation produced in radiation detectors [2]. Photoelectrons emitted into the gas medium undergo backscattering to the photocathode [3]; this effect, that depends on the wavelength, gas and electric field, reduces the transmission (photoelectron extraction efficiency) of the photoelectrons to the electron multiplier, lowering the effective extraction of the photoelectrons from the photocathode (effective quantum efficiency) as compared to vacuum. The effect is particularly significant in noble gases due to the high rate of elastic scattering at low electron energies. The addition of some molecular gases is known to increase transmission, when vibrational excitations of the molecules by electron impact competes efficiently with elastic scattering, cooling down the photoelectron energy in a few collisions to values that reduce the probability of their return to the photocathode (backscattering).

Photoelectron backscattering effects have been investigated experimentally by many authors and more recently by Monte Carlo simulation [3]-[9], examining the dependence of the photoelectron extraction efficiency $f$ on gas composition, photon energy $E_{ph}$ and electric field at the photocathode surface. Note that in earlier papers we have designated $f$ by electron *transmission* or *collection* efficiency, but in the present work we adopt the more common designation *extraction* efficiency.

Because the onset of multiplication is lower in Ne than in heavier noble gases like Ar or Xe, mixtures with Ne raise particular interest for micro patterned based gas devices, since higher gains can be reached at lower voltages [10]. Having this in mind, in the present work we investigated the photoelectron backscattering effects in mixtures of Ne with molecular gases: $CH_4$, $CF_4$, $CO_2$ and $N_2$. The experimental measurements are compared with Monte Carlo calculations, except for Ne-$N_2$.

## 2. Experimental set-up

The experimental set-up and the methodology for measuring photoelectron extraction efficiencies are described in detail in [7]. The photoelectron emission from CsI was induced by a Hg(Ar) VUV lamp (spectral distribution peaked in the VUV at 185 nm, with 5 nm *fwhm*). The chamber was pumped down to $10^{-5}$ Torr before being filled at 800 Torr and kept sealed; the relative uncertainty in the molecular gas concentrations in the mixtures is estimated not exceeding 0.1% of the total pressure. The extraction efficiency $f$ at a given condition was obtained as the ratio between the photocurrent values measured in the gas and in vacuum ($10^{-5}$ Torr) at equal electric fields.

## 3. Monte Carlo simulation

Detailed descriptions of the simulation model and the electron scattering cross-sections for Ne, $CH_4$, $CF_4$ and $CO_2$ used in the calculations can be found in [8], [9], [11]-[13], and the cross-sections are shown in figure 1. The scattering cross-sections by $N_2$ molecules are included in figure 1d and follow the data recommended in [14].

The emission of photoelectrons from the CsI photocathode was modelled in the calculations using the distribution for the photoelectron emission energy $\varepsilon_0$ measured in [15] for the photons from the VUV peak of the Hg(Ar) lamp. For each gas and field condition, the photoelectrons emitted from the photocathode were followed in the simulation along their free paths, taking into account elastic and inelastic collisions (and superelastic ones in $CF_4$ and $CO_2$). The extraction efficiency $f$ was calculated as the ratio $f = m/m_0$ between the number $m$ of photoelectrons not returning to the photocathode for a given number $m_0$ of photoelectrons (at least $2\times10^5$) emitted from the photocathode into the gas ($f = 1$ in vacuum).

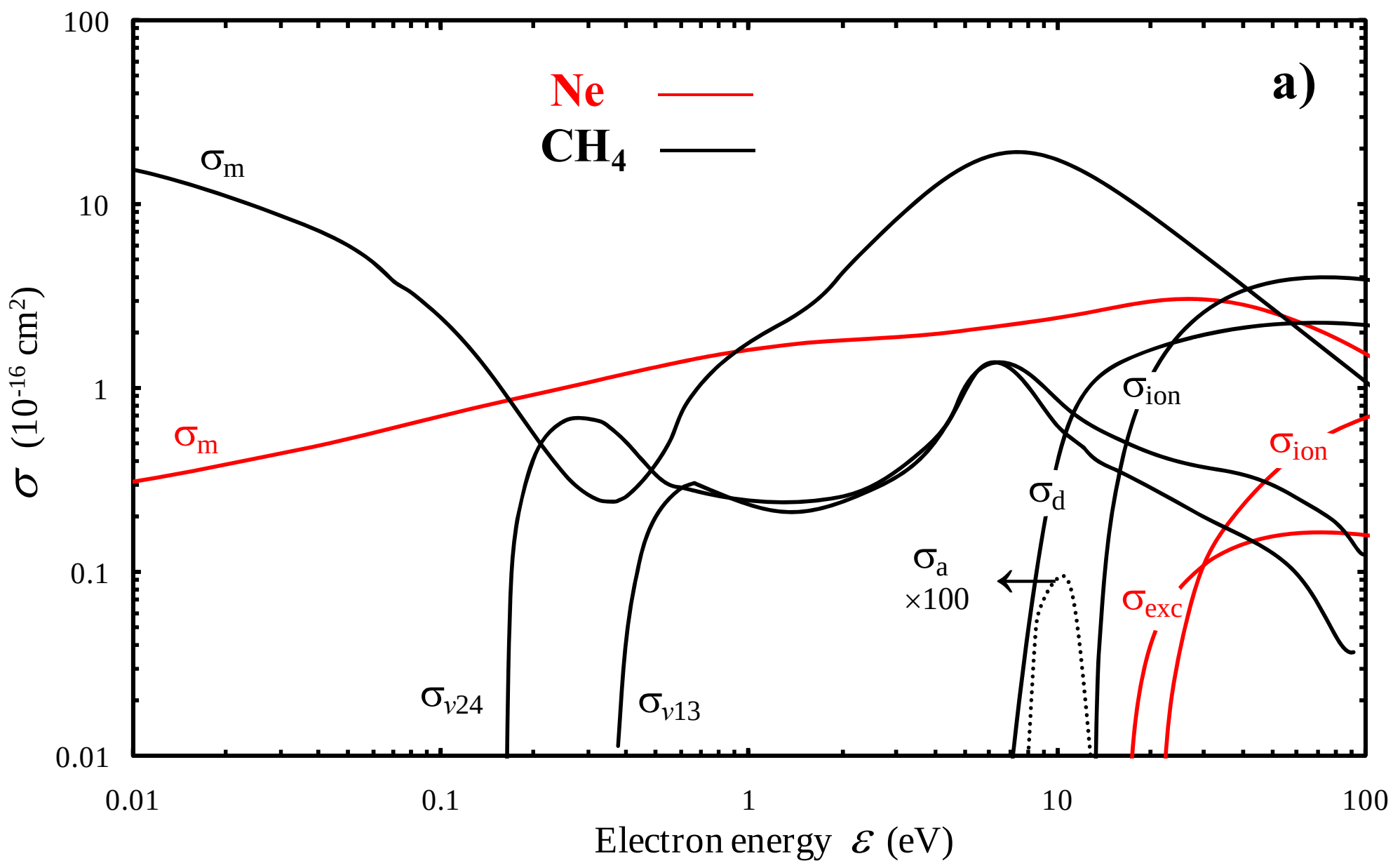

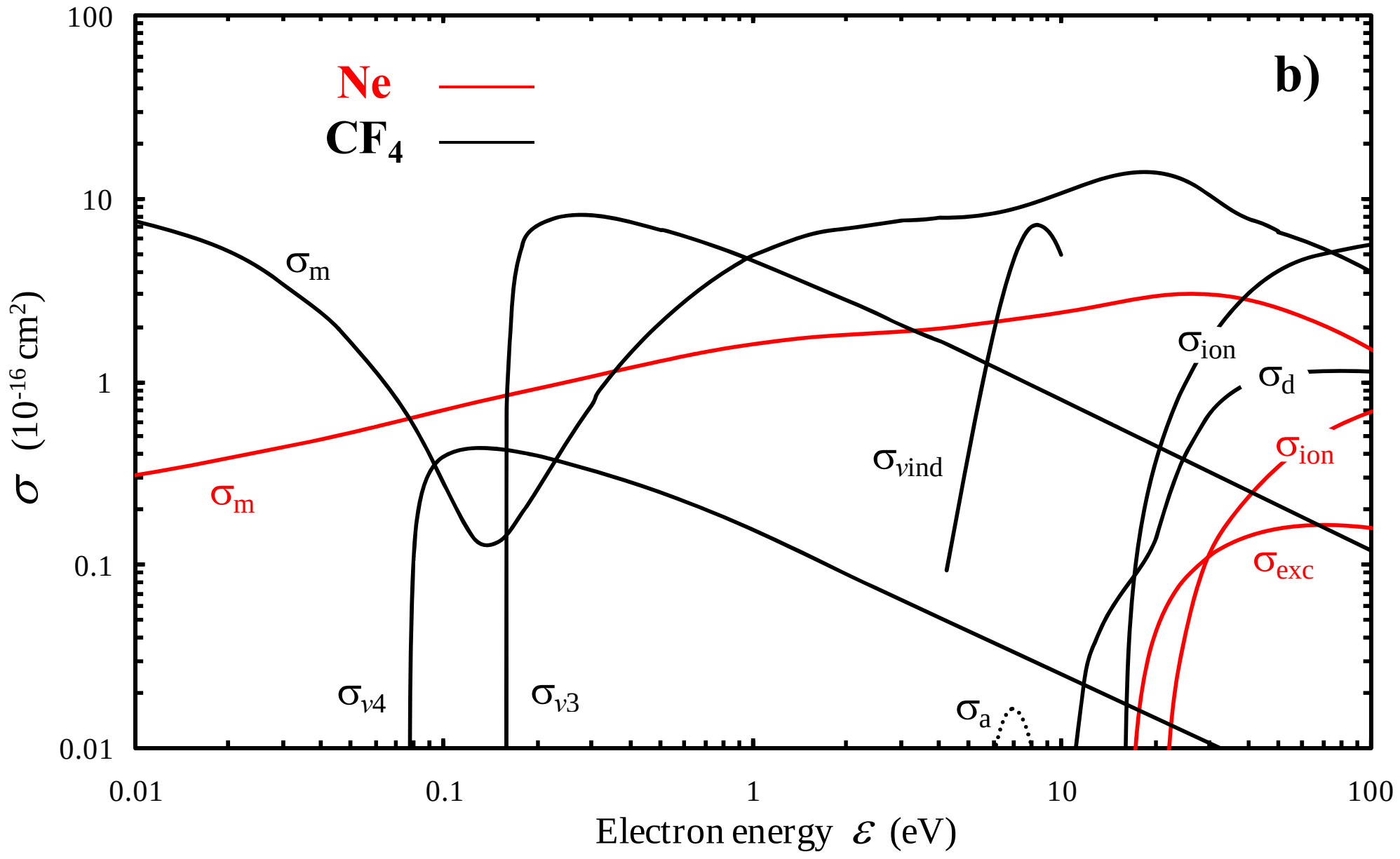

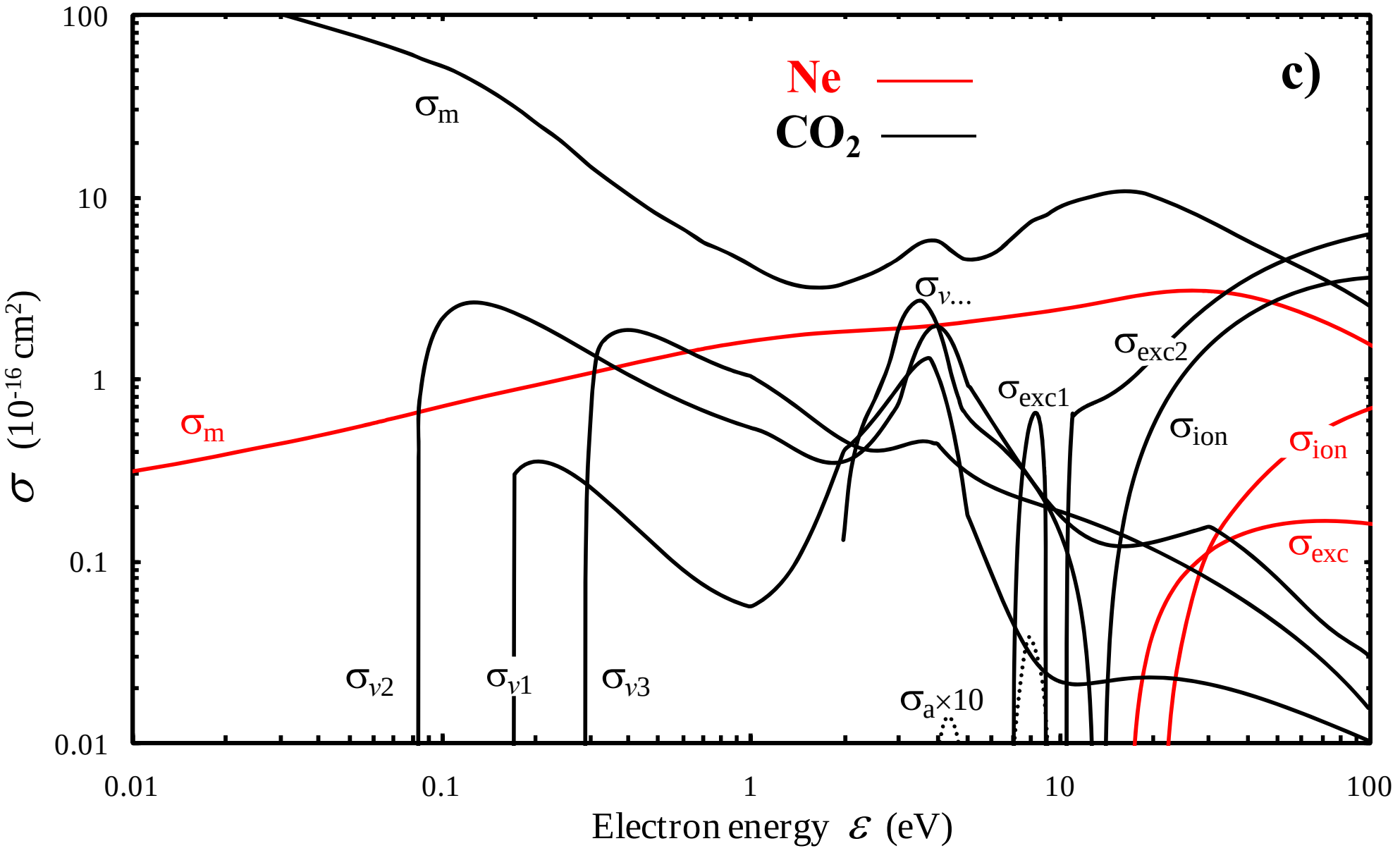

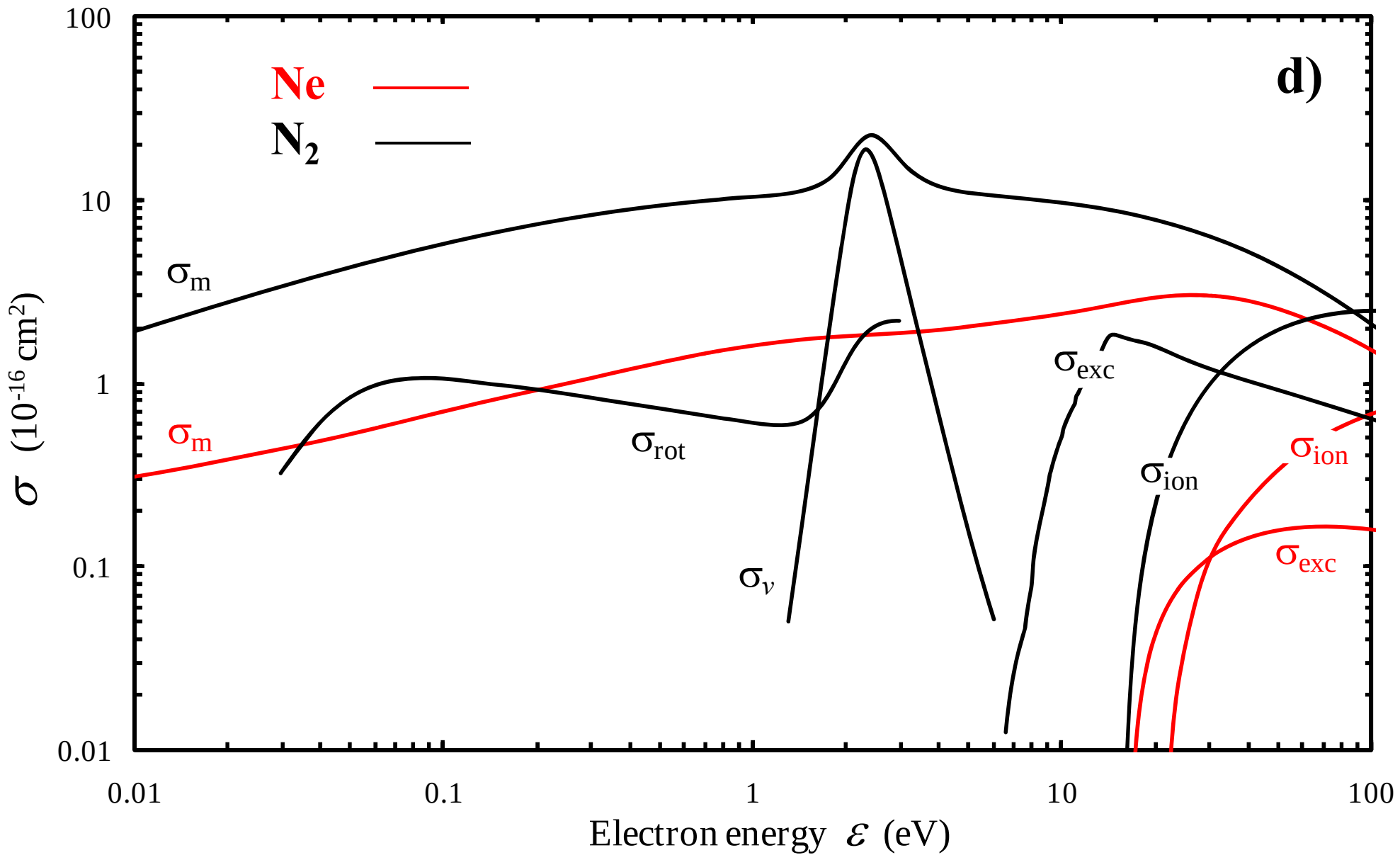

**Figure 1.** Electron scattering cross-sections in a) Ne and $CH_4$, b) Ne and $CF_4$, c) Ne and $CO_2$ d) Ne and $N_2$: elastic momentum transfer ($\sigma_m$), rotational excitation ($\sigma_{\text{rot}}$), vibrational excitation ($\sigma_v$), electron attachment ($\sigma_a$), neutral dissociation ($\sigma_d$), electronic excitation ($\sigma_{\text{exc}}$), and ionization ($\sigma_{\text{ion}}$).

## 4. Results and discussion

The measured extraction efficiencies $f$ are plotted in figure 2 as a function of the pressure-reduced electric field $E/p$ (and the electric field $E$ at 1 atm) for the mixtures investigated and in pure gases (Ne, $CH_4$, $CF_4$, $CO_2$ and $N_2$); the molecular gas concentrations were $\eta$ = 1%, 3%, 5%, 10%, 20%, 50% and 80%. In figure 3, the same data is represented as a function of the molecular gas concentration $\eta$ for the reduced electric field values $E/p$ = 0.1, 0.2, 0.3, 0.5, 0.8, 1, 2 and 3 V cm$^{-1}$ Torr$^{-1}$. The experimental $f$ values in figures 2 and 3 are compared with Monte Carlo simulation results, except for Ne-$N_2$. For practical comparative evaluation, figure 4 depicts the experimental $f$ curves for all Ne mixtures investigated with molecular gas concentrations $\eta$=10% and in the pure gases as a function of $E$ at 1 atm.

In figure 2 we observe that $f$ in the mixtures and in the pure gases increases, reflecting essentially the electron guiding at progressively higher electric fields. The increase tends to be faster at the lower-fields region. The effect of positive feedback caused by the absorption of the Ne self-scintillation photons by the photocathode is clearly visible in the experimental results in Ne and in its mixtures with low additive concentrations (reflected by a steep rise in the $f$ curves slope). This effect is not included in the simulations.

In figure 3, we see that $f$ increases always in Ne-$CH_4$ with the molecular concentration $\eta$; in Ne-$CF_4$ and Ne-$CO_2$ it increases until reaching a maximum and decreases afterwards towards a value in the corresponding pure molecular gas. The maximum is more pronounced, and the subsequent decrease is faster, in Ne-$CO_2$ than in Ne-$CF_4$. In the region of $\eta$ <~5%, in Ne-$CH_4$, Ne-$CF_4$ and Ne-$CO_2$ $f$ increases rapidly with $\eta$, and is thus expected to be very sensitive to variations in $\eta$ in this range. A different behaviour is observed in Ne-$N_2$, where the $f$ curves mostly decrease with $N_2$ concentration. The results are discussed below.

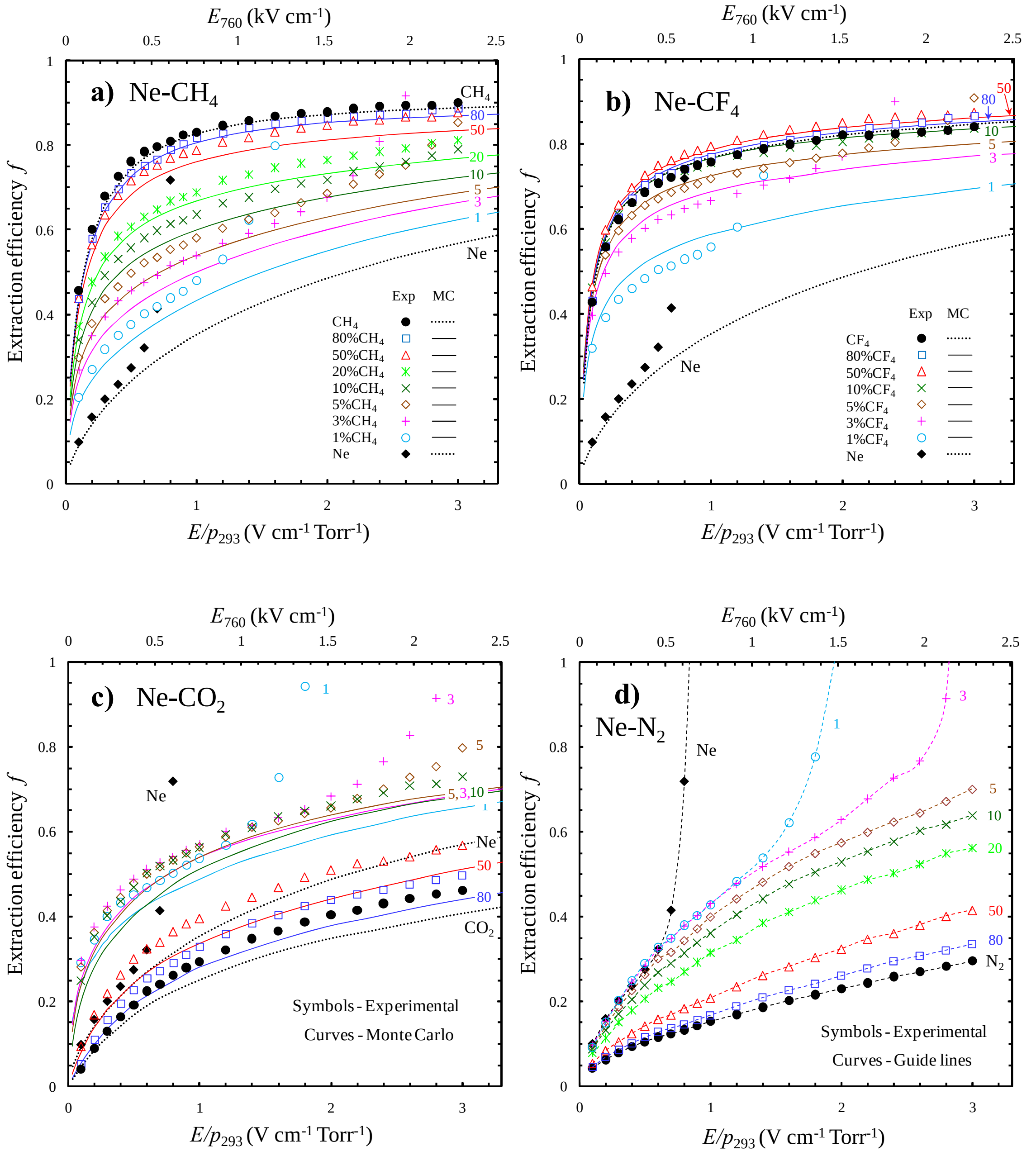

**Figure 2.** Experimental (symbols) and Monte Carlo (full and dotted curves) results of the photoelectron extraction efficiency $f$ from CsI as a function of the reduced electric field $E/p$ (bottom scale) and electric field at 1 atm (top scale) for a) Ne-$CH_4$ b) Ne-$CF_4$ c) Ne-$CO_2$ and d) Ne-$N_2$; data for Ne, $CH_4$, $CF_4$, $CO_2$ and $N_2$ are given as well; the concentrations $\eta$ (%) of the molecular gas are indicated (similar symbols for all figures). Note that in d) simulation results were not available for Ne-$N_2$; the dashed curves are just guidelines. The effect of positive feedback from Ne self-scintillation is visible only in the experimental results in mixtures with lower $\eta$-values. The Monte Carlo results in Ne and Ne-$CH_4$ are from [11] and in $CH_4$ from [9]; the measurements in $CH_4$ are from [7].

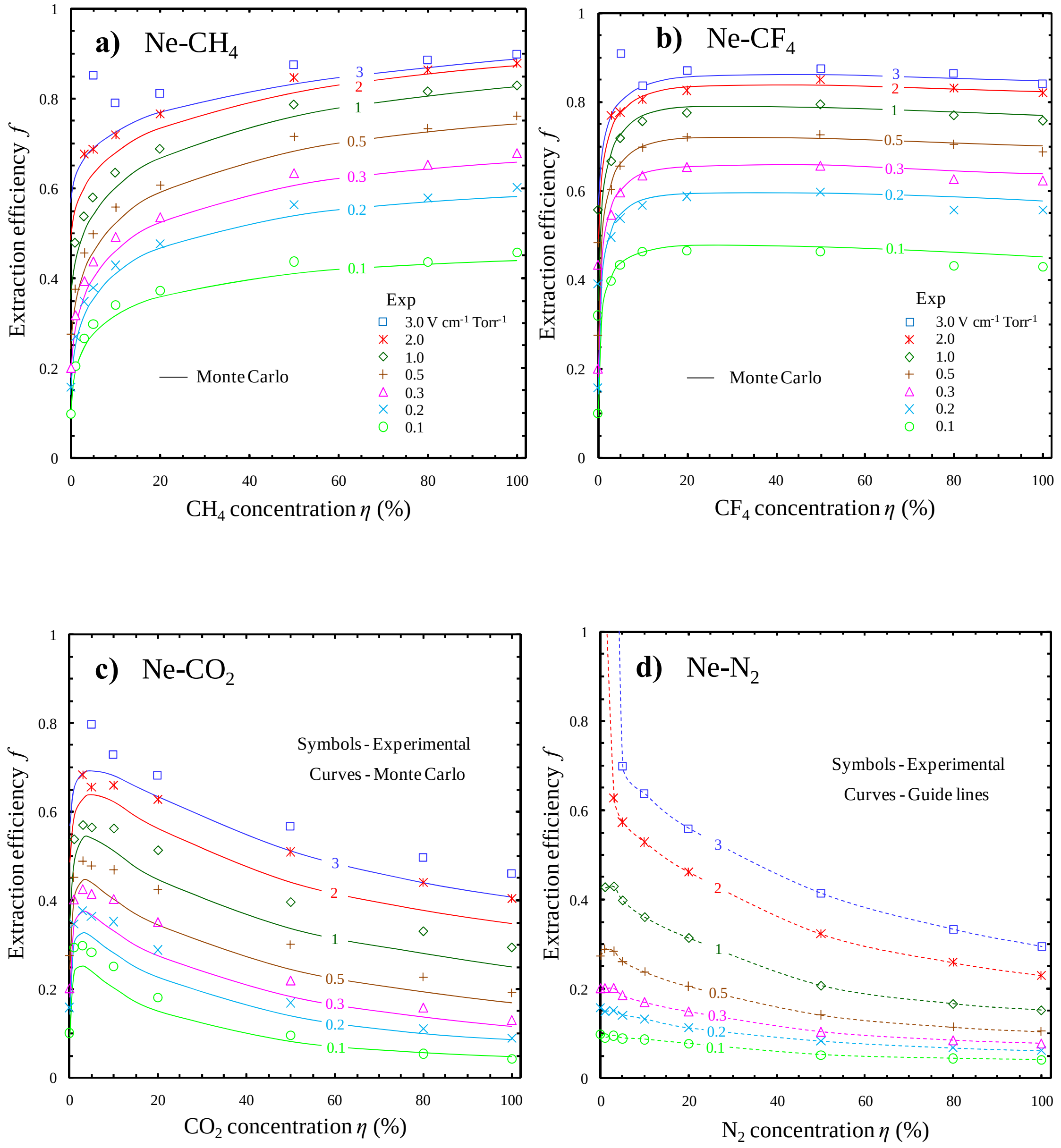

**Figure 3.** Experimental (symbols) and Monte Carlo (full curves) results of the photoelectron extraction efficiency *f* from CsI into a) Ne-$CH_4$ b) Ne-$CF_4$ c) Ne-$CO_2$ and d) Ne-$N_2$ mixtures, as a function of the concentration $\eta$ of the molecular gas; the values of the reduced electric field $E/p$ are indicated. (symbols refer to the same concentrations in all figures). Note that in d) the dashed curves are just guidelines. Deviations at low concentrations are due to positive feedback from Ne self-scintillation. The Monte Carlo results in Ne and Ne-$CH_4$ are from [11] and in $CH_4$ from [9]; the measurements in $CH_4$ are from [7].

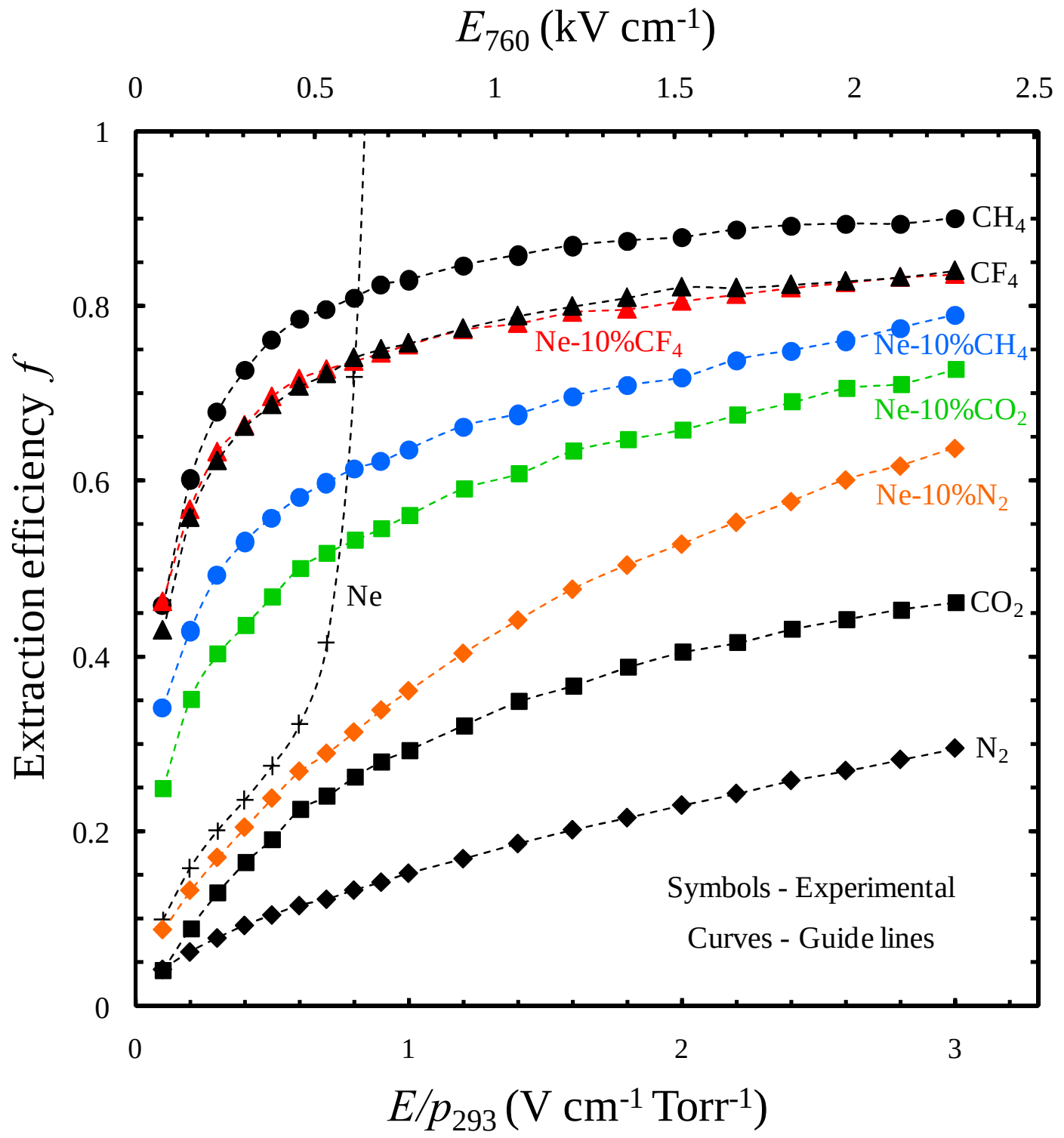

**Figure 4.** Summary of the experimental extraction efficiency $f$ from CsI as a function of reduced electric field (bottom scale) and electric field at 1 atm (top scale) into the Ne-$CH_4$, Ne-$CF_4$, Ne-$CO_2$ and Ne-$N_2$ mixtures with molecular gas concentration $\eta$ =10%, and into Ne, $CH_4$, $CF_4$, $CO_2$ and $N_2$. The effect of positive feedback from Ne self-scintillation is visible on the Ne curve. The curves joining the data points are guidelines.

We also observe in figure 3 that for Ne mixtures with $\eta$ = 5%-20% $f$ decreases as the additive gas varies from $CF_4$ to $CH_4$, to $CO_2$ and to $N_2$, while for higher $\eta$-values and in the pure gases $f$ decreases as we go from $CH_4$ to $CF_4$, to $CO_2$ and to $N_2$. Figure 4 shows that $f_{CH4}>f_{CF4}>f_{Ne}>f_{CO2}>f_{N2}$, except for the region above $E/p$~0.5 V cm$^{-1}$ Torr$^{-1}$ where positive feedback from Ne self-scintillation causes the Ne curve to rise steeply above all the molecular gases. The results are discussed below. Note that the result $f_{CH4}>f_{CF4}$ contradict [4], but more recent measurements in [16] confirm this sequence between the $f$–values in $CH_4$ and in $CF_4$.

Excluding the effect of positive feedback from Ne self scintillation, not included in the Monte Carlo simulations, there is good agreement between the experimental and the Monte Carlo results for $f$. However, the experimental values tend to be higher than the calculations. As discussed before [9], this may be attributed to a number of reasons, namely the reflection of the photoelectrons by the photocathode, uncertainty in the mixture composition, and uncertainty in the distribution of the photoelectron initial energies $\varepsilon_0$.

As discussed in [9], [11], the Monte Carlo simulations show that *f*-values in mixtures of noble gases with molecular additives are favoured not only by longer initial free paths of electrons in gas (governed by the predominant cross-section for elastic scattering), but also by a swift drop of the photoelectron initial energy through inelastic collisions with the molecules at low energies. Accordingly, the behaviour of *f* is found to be related to the ratios between the number of vibrational excitation collisions $n_v$ and the total number $n_t$ of collisions in the gas: $\varsigma = n_v/n_t$, $\varsigma$ being a measure of the competition between vibrational excitation of the molecules and elastic scattering.

In the present measurements, where the CsI photocathode was irradiated by $\lambda$=185 nm ($E_{ph}$=6.7 eV) photons, most of the photoelectrons are emitted at initial energies $\varepsilon_0$ in the range 0.1-0.5 eV [15]. Therefore, the addition of a molecular gas with low cross-section for elastic scattering and with high enough $\varsigma$-values in the photoelectron energy range is expected to improve photoelectron transmission. $CH_4$, $CF_4$ and $CO_2$ fulfil this requirement, since they exhibit large vibrational excitation cross-sections in the energy range of the photoelectrons, and moreover the elastic cross-sections in $CH_4$ and $CF_4$ exhibit pronounced Ramsauer-Townsend minima in the same region. Accordingly, the measurements and the simulation results (figures 2 to 4) show that the addition of small amounts of these molecular gases to Ne significantly improve *f*, because the initial photoelectron energy may be reduced after a few collisions to values that cut down the chance of return to the photocathode.

The behaviour of *f* with $\eta$ and *E*/*p* in figures 2 to 4 for different mixtures depends on the behaviour with energy of the involved elastic and vibrational excitation scattering cross-sections (figure 1), the degree of their overlap and the energy which is lost in vibrational excitation of the molecules; but, it is clear that the additives $CH_4$ and $CF_4$ cause in general a better improvement of *f* in Ne mixtures than $CO_2$ does. This happens essentially because as we move to higher $\eta$-values the vibrational-to-elastic ratios $\varsigma$ decrease and become progressively lower for Ne-$CO_2$ and also because the addition of $CO_2$ always causes a larger increase of the photoelectrons elastic scattering (decrease of penetration) than $CH_4$ or $CF_4$ do. Accordingly, *f* in pure $CO_2$ is much lower than in pure $CH_4$ or $CF_4$.

Comparing the *f* curves in pure $CH_4$ and $CF_4$, we observe that *f* tends to be higher in $CH_4$ than in $CF_4$, see figure 4 in particular. The simulations show that this occurs because, for the present range of photoelectron energies, the penetrations are higher in $CH_4$ than in $CF_4$, and, even though $\varsigma$ in $CH_4$ is lower, electrons loose on the average more energy in vibrational excitation collisions with the $CH_4$ molecules. Naturally, *f* in Ne-$CH_4$ mixtures is also larger than in Ne-$CF_4$ for high $\eta$-values, but as we move towards lower $\eta$-values the order is reversed (*f* in Ne-$CF_4$ becomes larger than in Ne-$CH_4$; compare figure 3a with 3b and the $\eta$=10% curves in figure 4). This is because at decreasing $\eta$, the photoelectron penetrations in Ne-$CF_4$ increase and approach those in Ne-$CH_4$ and $\varsigma$ in Ne-$CF_4$ becomes much larger than in Ne-$CH_4$.

Finally, in what concerns $N_2$, the measurements in figures 2 to 4 have shown firstly that no improvement of *f* is gained by adding $N_2$ to Ne, and secondly that the *f* values in $N_2$ are lower than in $CH_4$, $CF_4$ or $CO_2$. This behaviour is consistent with the low photoelectron penetrations associated to the high cross-section for elastic scattering by $N_2$ molecules (see figure 1d) and the fact that the range where the vibrational excitation of $N_2$ is significant (above ~1 eV) falls beyond the photoelectron energy range in the present investigation. Monte Carlo calculations are still not available for $N_2$ and Ne-$N_2$ mixtures, but the calculations are expected to reproduce the behaviour of the experimental data.

## 5. Conclusions

In this work we have investigated electron backscattering effects and have measured the extraction efficiency $f$ for the photoelectrons emitted from CsI into mixtures of Ne with molecular gases $CH_4$, $CF_4$, $CO_2$ and $N_2$ when the photocathode is irradiated with UV photons of $\lambda$=185 nm ($E_{ph}$=6.7 eV); most photoelectrons had initial energies $\varepsilon_0$ in the range ~ 0.1 to 0.5 eV. The extraction efficiencies $f$ were measured at reduced electric fields below 3 V $cm^{-1}$ $Torr^{-1}$ (E<2.4 kV cm-1 at 1 atm); the results were compared with Monte Carlo simulations, except for Ne-$N_2$.

In the conditions of the present experiments (for the given $E_{ph}$ and $\varepsilon_0$ range) it was shown that the addition of $CH_4$, $CF_4$ or $CO_2$ to Ne significantly reduced photoelectron backscattering, enhancing photoelectron extraction. It was found that $CH_4$ or $CF_4$ are in general more efficient additives than $CO_2$. The efficiency $f$ in Ne-$CF_4$ tends to be higher than in Ne-$CH_4$, at low additive concentrations; the order is reversed at higher concentrations $\eta$ of the molecular gas, resulting in $f_{CF4} < f_{CH4}$ in pure $CH_4$ and $CF_4$.

On the other hand, the measurements indicate that the addition of $N_2$ to Ne does not improve $f$, because the elastic scattering cross-section by $N_2$ molecules is high and the range where the vibrational excitation is significant falls beyond the photoelectron energy range in the present work.

The results of this work may have an impact on the choice of operation gas in UV gaseous detectors for RICH. Recent results on high-gain operation of CsI-coated THGEM UV-detectors in Ne/$CH_4$ and Ne/$CF_4$ are published in [16].

## Acknowledgments

This work was carried out at Centro de Instrumentação (Research Unit 217/94), Physics Department, University of Coimbra, Portugal, and was supported by FEDER and POCI2010 programs through FCT (Fundação para a Ciência e Tecnologia, Portugal) projects POCI/FP/81935/2007 and PTDC/FIS/100474/2008. J. Escada acknowledges support from FCT Grant SFRH/BD/22177/2005. A. Breskin is the W. P. Reuther Professor of Research in the peaceful use of Atomic Energy.